\documentclass[trackchanges, twocolumn, twocolappendix]{aastex701}

\usepackage{multirow, amsmath}
\usepackage{dsfont}
\usepackage{comment}

\begin{document}

\title{Joint Estimation of Properties of the Lunar Subsurface and Galactic Foregrounds with LuSEE-Night}

\author[orcid=0000-0002-3152-796X, sname='Yousuf']{Fatima Yousuf}
\affiliation{Department of Physics, University of California, Berkeley, CA, 94720-7300, USA}
\affiliation{Space Sciences Laboratory, University of California, Berkeley, CA 94720-7450, USA}
\email[show]{fatima\_yousuf@berkeley.edu}

\author[orcid=0000-0002-0309-9750, sname='Li']{Zack Li}
\affiliation{Berkeley Center for Cosmological Physics, University of California, Berkeley, CA 94720, USA}
\affiliation{Physics Division, Lawrence Berkeley National Laboratory, Berkeley, CA 94720, USA}
\email{zackli@berkeley.edu}

\author[orcid=0000-0002-1989-3596, sname='Bale']{Stuart D. Bale}
\affiliation{Department of Physics, University of California, Berkeley, CA, 94720-7300, USA}
\affiliation{Space Sciences Laboratory, University of California, Berkeley, CA 94720-7450, USA}
\affiliation{The Blackett Laboratory, Imperial College London, London, SW7 2AZ, UK}
\email{bale@berkeley.edu}

\author[orcid=0009-0002-4218-4840, sname='Barker']{David W. Barker}
\affiliation{Center for Astrophysics and Space Astronomy, Department of Astrophysical and Planetary Sciences, University of Colorado Boulder, CO 80309, USA}
\email{David.W.Barker@Colorado.EDU}

\author[orcid=0000-0002-4468-2117, sname='Burns']{Jack Burns}
\affiliation{Center for Astrophysics and Space Astronomy, Department of Astrophysical and Planetary Sciences, University of Colorado Boulder, CO 80309, USA}
\email{Jack.Burns@colorado.edu}

\author[orcid=0000-0002-7971-3390, sname='Bye']{Christian H. Bye}
\affiliation{Department of Astronomy, University of California, Berkeley, CA 94720, USA}
\affiliation{Radio Astronomy Laboratory, University of California, Berkeley, CA 94720, USA}
\email{chbye@berkeley.edu}

\author[orcid=0000-0001-5871-0951, sname='Camacho']{Hugo Camacho}
\affiliation{Brookhaven National Laboratory, Upton, NY 11973, USA}
\email{hcamachoc@bnl.gov}

\author[orcid=0009-0003-5974-0185, sname='Cordun']{Cristina-Maria Cordun}
\affiliation{ASTRON, Netherlands Institute for Radio Astronomy, Oude Hoogeveensedĳk 4, Dwingeloo, 7991 PD, The Netherlands}
\email{cordun@astron.nl}

\author[orcid=0000-0002-3292-9784, sname='Dorigo Jones']{Johnny Dorigo Jones}
\affiliation{Center for Astrophysics and Space Astronomy, Department of Astrophysical and Planetary Sciences, University of Colorado Boulder, CO 80309, USA}
\email{Johnny.DorigoJones@colorado.edu}

\author[orcid=0000-0002-7491-2753, sname='Fahs']{Adam Fahs}
\affiliation{Department of Physics, University of California, Berkeley, CA, 94720-7300, USA}
\affiliation{Physics Division, Lawrence Berkeley National Laboratory, Berkeley, CA 94720, USA}
\email{adam.fahs@berkeley.edu}

\author[orcid=0000-0002-6524-9830, sname='Ghosh']{Sonia Ghosh}
\affiliation{Kapteyn Astronomical Institute, University of Groningen, P.O.Box 800, 9700 AV Groningen, The Netherlands}
\email{soniaghosh@astro.rug.nl}

\author[orcid=0000-0003-0420-3633, sname='Goetz']{Keith Goetz}
\affiliation{School of Physics and Astronomy, University of Minnesota, Minneapolis, MN 55455, USA}
\email{goetz@umn.edu}

\author[orcid=0000-0002-7588-1194, sname='Grimm']{Robert Grimm}
\affiliation{Solar System Exploration Division, Southwest Research Institute, 1301 Walnut St, Suite 400, Boulder, CO 80302, USA}
\email{robert.grimm@swri.org}

\author[sname='Herrmann']{Sven Herrmann}
\affiliation{Brookhaven National Laboratory, Upton, NY 11973, USA}
\email{sherrmann@bnl.gov}

\author[orcid=0000-0002-9377-5133, sname='Hibbard']{Joshua J. Hibbard}
\affiliation{Space Sciences Laboratory, University of California, Berkeley, CA 94720-7450, USA}
\email{hibbardjj@berkeley.edu}

\author[orcid=0000-0001-8101-468X, sname='Jeong']{Oliver Jeong}
\affiliation{CNRS-UCB International Research Laboratory, Centre Pierre Bin\'etruy, IRL 2007, CPB-IN2P3, Berkeley, CA 94720, US}
\affiliation{Physics Division, Lawrence Berkeley National Laboratory, Berkeley, CA 94720, USA}
\email{objeong@lbl.gov}

\author[orcid=0000-0001-7901-9545, sname='Klein-Wolt']{Marc Klein-Wolt}
\affiliation{Department of Astrophysics, Research Institute of Mathematics, Astrophysics and Particle Physics, Radboud University Nijmegen, Heijendaalseweg 135, 6525 AJ Nijmegen, The Netherlands}
\email{M.KleinWolt@astro.ru.nl}

\author[orcid=0000-0003-1840-0312, sname='Koopmans']{L\'{e}on V.E. Koopmans}
\affiliation{Kapteyn Astronomical Institute, University of Groningen, P.O.Box 800, 9700 AV Groningen, The Netherlands}
\email{koopmans@astro.rug.nl}

\author[sname='Krajewski']{Joel Krajewski}
\affiliation{Space Sciences Laboratory, University of California, Berkeley, CA 94720-7450, USA}
\email{joelkrajewski@berkeley.edu}

\author[orcid=0000-0002-9552-8822, sname='Louis']{Corentin Louis}
\affiliation{LIRA, Observatoire de Paris, Université PSL, Sorbonne Université, Université Paris Cité, CY Cergy Paris Université, CNRS, 92190 Meudon, France}
\email{corentin.louis@obspm.fr}

\author[orcid=0000-0001-6172-5062, sname='Maksimović']{Milan Maksimovi\'{c}}
\affiliation{LIRA, Observatoire de Paris, Université PSL, Sorbonne Université, Université Paris Cité, CY Cergy Paris Université, CNRS, 92190 Meudon, France}
\email{milan.maksimovic@obspm.fr}

\author[orcid=0009-0006-6700-5692, sname='McLean']{Ryan McLean}
\affiliation{Space Sciences Laboratory, University of California, Berkeley, CA 94720-7450, USA}
\email{rmclean@ssl.berkeley.edu}

\author[orcid=0000-0002-3287-2327, sname='Monsalve']{Raul A. Monsalve}
\affiliation{Space Sciences Laboratory, University of California, Berkeley, CA 94720-7450, USA}
\affiliation{School of Earth and Space Exploration, Arizona State University, Tempe, AZ 85287, USA}
\affiliation{Departamento de Ingeniería Eléctrica, Universidad Católica de la Santísima Concepción, Alonso de Ribera 2850, Concepción, Chile}
\email{raul.monsalve@berkeley.edu}

\author[orcid=0000-0003-4823-5311, sname='Nigmetov']{Arnur Nigmetov}
\affiliation{Physics Division, Lawrence Berkeley National Laboratory, Berkeley, CA 94720, USA}
\email{anigmetov@berkeley.edu}

\author[orcid=0000-0002-8718-2235, sname='O'Connor']{Paul O'Connor}
\affiliation{Brookhaven National Laboratory, Upton, NY 11973, USA}
\email{poc@bnl.gov}

\author[orcid=0000-0002-5400-8097, sname='Parsons']{Aaron Parsons}
\affiliation{Department of Astronomy, University of California, Berkeley, CA 94720, USA}
\affiliation{Radio Astronomy Laboratory, University of California, Berkeley, CA 94720, USA}
\email{aparsons@berkeley.edu}

\author[orcid=0000-0003-0643-3088, sname='Piat']{Michel Piat}
\affiliation{Laboratoire Astroparticule et Cosmologie (APC), Université Paris-Cité, Paris, France}
\email{piat@apc.in2p3.fr}

\author[orcid=0000-0002-1573-7457, sname='Pulupa']{Marc Pulupa}
\affiliation{Space Sciences Laboratory, University of California, Berkeley, CA 94720-7450, USA}
\email{pulupa@berkeley.edu}

\author[orcid=0009-0001-0044-4600, sname='Pund']{Rugved Pund}
\affiliation{Physics and Astronomy Department, Stony Brook University, Stony Brook, NY 11794, USA}
\affiliation{Brookhaven National Laboratory, Upton, NY 11973, USA}
\email{rugved.pund@stonybrook.edu}

\author[orcid=0000-0003-2196-6675, sname='Rapetti']{David Rapetti}
\affiliation{NASA Ames Research Center, Moffett Field, CA 94035, USA}
\affiliation{Research Institute for Advanced Computer Science, Universities Space Research Association, Washington, DC 20024, USA}
\affiliation{Center for Astrophysics and Space Astronomy, Department of Astrophysical and Planetary Sciences, University of Colorado Boulder, CO 80309, USA}
\email{David.Rapetti@Colorado.EDU}

\author[orcid=0000-0002-9181-9948, sname='Rotermund']{Kaja M. Rotermund}
\affiliation{Physics Division, Lawrence Berkeley National Laboratory, Berkeley, CA 94720, USA}
\email{KRotermund@lbl.gov}

\author[orcid=0000-0002-5089-7472, sname='Saliwanchik']{Benjamin Saliwanchik}
\affiliation{Brookhaven National Laboratory, Upton, NY 11973, USA}
\email{bsaliwanc@bnl.gov}

\author[orcid=0000-0002-8713-3695, sname='Slosar']{An\v{z}e Slosar}
\affiliation{Brookhaven National Laboratory, Upton, NY 11973, USA}
\email{anze@bnl.gov}

\author[sname='Speedie']{Graham Speedie}
\affiliation{Physics and Astronomy Department, Stony Brook University, Stony Brook, NY 11794, USA}
\affiliation{Brookhaven National Laboratory, Upton, NY 11973, USA}
\email{graham.speedie@stonybrook.edu}

\author[sname='Stefanov']{Nikolai Stefanov}
\affiliation{Department of Physics, University of California, Berkeley, CA, 94720-7300, USA}
\affiliation{Space Sciences Laboratory, University of California, Berkeley, CA 94720-7450, USA}
\email{stefanov@berkeley.edu}

\author[orcid=0000-0003-2794-7926, sname='Sundkvist']{David Sundkvist}
\affiliation{Space Sciences Laboratory, University of California, Berkeley, CA 94720-7450, USA}
\email{sundkvistd@berkeley.edu}

\author[orcid=0000-0001-8101-468X, sname='Suzuki']{Aritoki Suzuki}
\affiliation{Physics Division, Lawrence Berkeley National Laboratory, Berkeley, CA 94720, USA}
\email{asuzuki@lbl.gov}

\author[orcid=0000-0002-0872-181X, sname='Vedantham']{Harish K. Vedantham}
\affiliation{ASTRON, Netherlands Institute for Radio Astronomy, Oude Hoogeveensedĳk 4, Dwingeloo, 7991 PD, The Netherlands}
\affiliation{Kapteyn Astronomical Institute, University of Groningen, P.O.Box 800, 9700 AV Groningen, The Netherlands}
\email{vedantham@astron.nl}

\author[orcid=0000-0003-1672-9878, sname='Zarka']{Philippe Zarka}
\affiliation{LIRA, Observatoire de Paris, Université PSL, Sorbonne Université, Université Paris Cité, CY Cergy Paris Université, CNRS, 92190 Meudon, France}
\email{philippe.zarka@obspm.fr}
\collaboration{all}{The LuSEE-Night Collaboration}

\begin{abstract}

The Lunar Surface Electromagnetics Experiment (LuSEE-Night) is a joint NASA-DOE-ESA low-frequency radio telescope that will reach the lunar far side in 2027.  The unknown dielectric properties of the subsurface at the LuSEE-Night landing site impose the most significant limitation for precision instrument calibration, as reflections from the lunar subsurface can change the primary beam at the 10-20\% level. Simulations of these effects have provided insight and concern, showing that the lunar subsurface modeled as a lossy dielectric can absorb a large amount of the power of the sky signal. While this absorption may not strongly impact the signal-to-noise ratio in a sky noise-dominated regime, it could complicate the beam pattern and make the signal more difficult to model and interpret. We have simulated the far-field properties of the LuSEE-Night beam for varying dielectric profiles of the lunar subsurface. We find that varying the properties of the lunar subsurface has the most significant impact around the antenna resonance, impacting its amplitude, position and width. Conversely, changing the properties of the foreground impacts the data across the band.  We use a Bayesian inference pipeline to jointly estimate parameters of a Galactic foreground model and dielectric properties of the lunar subsurface around the LuSEE-Night landing site and find that parameters of both the galaxy and subsurface properties can be estimated jointly.  While the modeling is somewhat idealized, we believe that the results are largely robust owing to the fact that spectral variations for plausible subsurface and galaxy models have very different spectral signatures.
\end{abstract}

\keywords{\uat{Cosmology}{343} --- \uat{Radio receivers}{1355} --- \uat{Bayesian Statistics}{1900} --- \uat{H I line emission}{690} }


\section{Introduction} 

The 21-cm signal of neutral hydrogen is predicted to be a sensitive probe of the universe at the Epoch of Reionization (EoR) ($z \sim 6 - 15$) and Cosmic Dawn ($z \sim 15 - 30$), and the sole probe of the Dark Ages ($z \sim 30 - 200$) \citep{Furlanetto2006-dp}. Measurements of the Dark Ages 21-cm signal could give information about neutrino decay lifetimes, measure the effect of gravitational waves on large scale structures, and further constrain $\Lambda$CDM cosmological models \citep{mondal2023,CHIANESE201964,mishra2018}. 

Observations from the lunar farside provide a way of probing the state of neutral hydrogen during the Dark Ages while bypassing the issues of terrestrial ionospheric distortions and radio frequency interference (RFI) from Earth. However, the Dark Ages spectral feature in particular is still challenging to recover since it is intrinsically faint (predicted to be $\sim$ tens of mK) and more than 5 orders of magnitude below the Galactic foreground \citep{Mondal2023-rt}. Moreover, the Dark Ages 21-cm signal redshifts to frequencies between 3-30 MHz, which are especially difficult to study from Earth. This is due to the $\sim$10 MHz plasma frequency of the F-layer peak of the terrestrial ionosphere, below which the ionosphere becomes opaque and radio waves below $\sim$300 MHz are refracted and partially absorbed \citep{koopmans2016}. 

The Cosmic Dawn and EoR 21-cm signals are a prime target for multiple ground-based radio interferometers, including the Murchison Widefield Array (MWA) \citep{Tingay2013-di}, the Hydrogen Epoch of Reionization Array (HERA) \citep{DeBoer2017-oo}, the Low Frequency ARray (LOFAR) \citep{Van_Haarlem2013-mz}, the New extension in Nan{\c c}ay upgrading LOFAR (NenuFAR) \citep{Zarka2018-cr, Munshi2024-vs, nenufar2025}, the Precision Array to Probe the Epoch of Reionization (PAPER) \citep{Parsons2010-kh}, and the Giant Meter-wave Radio Telescope (GMRT) \citep{gmrt}. Additionally, there are several proposals for radio telescopes on the lunar farside, including FarView \citep{burns2021lunarfarsidelowradio, burns2026farview}, DEX \citep{brinkerink2025darkagesexplorerdex}, the Lunar Crater Radio Telescope (LCRT) \citep{lcrt_Bandyopadhyay}, the Large-scale Array for Radio Astronomy on the Farside (LARAF) \cite{Chen_2024}, and the Lunar Farside Technosignature and Transient Telescope (LFT3) \citep{DeBoer2025}. Proposals for lunar-orbiting observatories include the Probing ReionizATion of the Universe using Signal from Hydrogen (PRATUSH) experiment \citep{Sathyanarayana_Rao2023} and the Discovering the Sky at the Longest Wavelengths (DSL/Hongmeng) constellation \citep{chen2020}, which aim to exploit the radio-quiet environment of the lunar farside from orbit. 

Lunar surface-based instruments like LuSEE-Night tend to be more expensive than lunar orbital observatories, and they run a greater risk of crashing into the lunar surface. The subsurface structure also imposes complexity to beam modelling \citep{Burns2021-md}.  However, unlike orbital observatories, which observe the lunar farside only during a fraction of each orbit, lunar surface-based instruments such as LuSEE-Night can integrate continuously over a full lunar night ($\sim$2 weeks). During this time the instrument can settle into an extremely temperature stable and repeatable state, while orbiters must contend with rapid temperature changes caused by a much shorter orbital period. Moreover, the fraction of time an orbiter spends in a truly shielded zone can be considerably less than 50\%, which lowers the effective integration speed.

LuSEE-Night is a pathfinder low frequency radio astronomy telescope that will be delivered to the lunar far side by the NASA Commercial Lunar Payload Services program in 2027. The payload system has been developed jointly by NASA, the US Department of Energy, and the UC Berkeley Space Sciences Laboratory. The European Space Agency (ESA) provides the capabilities for sending data back to Earth via the Lunar Pathfinder (LPF). The LuSEE-Night instrument includes four 3m BeCu antennas configured in orthogonal dipole pairs. The antennas are mounted on a motor-driven turntable so that the antenna orientation can rotate in the plane of the lunar surface. The antennas are sensitive to linear polarizations, providing additional information that helps distinguish the unpolarized Dark Ages 21-cm signal from partially polarized Galactic foregrounds. A preliminary description of the instrument can be found in \cite{bale2023luseenightlunarsurface}, with a more complete instrument design paper in preparation.

\begin{figure}
\includegraphics[width=0.9\linewidth]{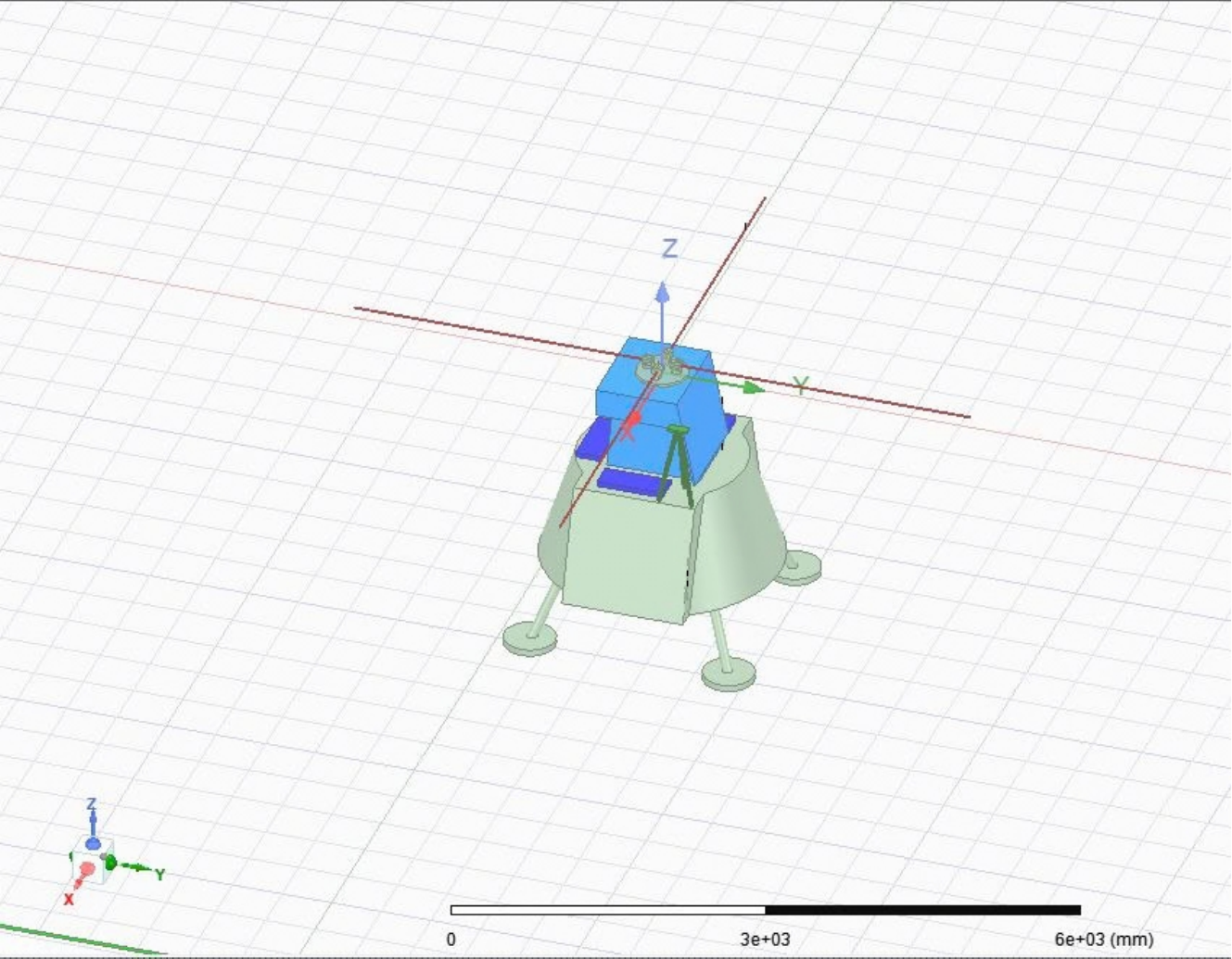}
\centering
\caption{\label{fig:figure1}Model of LuSEE-Night instrument and lander in Ansys HFSS used to simulate antenna far-field beam pattern. The instrument includes four 3m BeCu stacer antenna monopole elements arranged in a cross-dipole configuration.}
\end{figure}

\begin{figure}
\includegraphics[width=\linewidth]{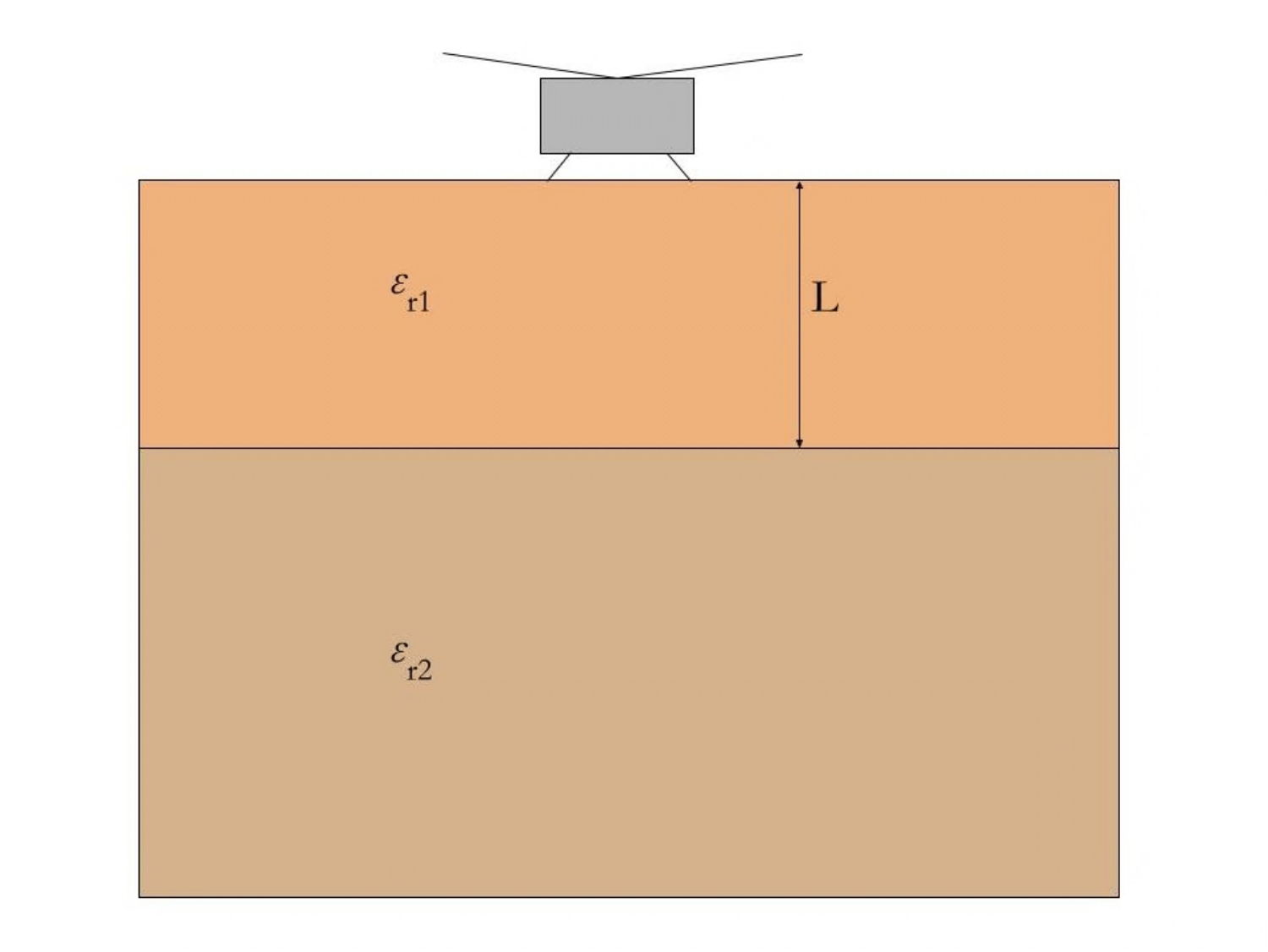}
\centering
\caption{\label{fig:figure2}In HFSS, the lunar subsurface is modeled as a layered impedance boundary. The top layer has a depth of $L$ and dielectric constant of $\epsilon_{r1}$. The bottom layer is infinitely deep and has a dielectric constant of $\epsilon_{r2}$. Modeling the lunar subsurface as a multi-layered, stratified medium is consistent with past work, including \cite{Fa2007-qi}, \cite{Chen2013-zu}, and \cite{GRIMM2018389}.}
\label{fig:regolith_structure}
\end{figure}

The radiation pattern of wide-beam antennas can be significantly influenced by the electromagnetic properties of the ground \citep{tokarsky2022, Tokarsky2023}. Several prior experiments have studied the lunar surface and demonstrated its transparency to radio waves, including the Apollo 17 Lunar Sounder Experiment \citep{phillipsApolloLunar}, the Apollo 17 Surface Electrical Properties Experiment \citep{Simmons_SEP,GRIMM2018389}, the Kaguya Lunar Radar Sounder \citep{Ono2009} and the Chang'e 3 and 4 rover Lunar Penetrating Radars \citep{Lai2019,Zhang2020}. 

Additionally, several ground-based 21-cm experiments have also evaluated the effect of the dielectric and conductive properties of terrestrial soil on signal detection, including the Mapper of the IGM Spin Temperature (MIST) \citep{Monsalve2024,Hendricksen2026,Altamirano2025}, the Experiment to Detect the Global EoR Signature (EDGES) \citep{Mahesh2021-sj, Bowman2018}, the Large-aperture Experiment to detect the Dark Ages (LEDA) \citep{Spinelli2022}, the Shaped Antenna Measurement of the background Radio Spectrum (SARAS) \citep{Agrawal2024-az}, the Radio Experiment for the Analysis of Cosmic Hydrogen (REACH) \citep{Cumner2022-un}, the Remote H$_1$ eNvironment Observer (RHINO) \citep{Bull2025-gp}, the Probing Radio Intensity at high-Z from Marion (PRI$^{Z}$M) experiment \citep{Philip2019-hk}, and the Electromagnetically Isolated Global Signal Estimation Platform (EIGSEP) \citep{Bye2026-hv}. 

\begin{table}
\centering
\caption{Lunar Subsurface Parameters \label{tab:regolith_params}}
\hspace{-2cm}
\begin{tabular}{ |c|c|c| } 
\hline
$\epsilon_{r,top}$ & $\epsilon_{r, bottom}$ & $L$ \\
\hline
3.2 & 3.8 & 0.5m \\ 
3.4 & 4.0 & 1.0m \\
3.6 & 4.2 & 1.5m \\
3.8 & 4.4 & 2.0m \\
4.0 & 4.6 & 2.5m \\
4.2 & 4.8 & 3.0m \\
\hline
\end{tabular}
\tablecomments{Parameter values used to simulate the lunar subsurface (see Figure~\ref{fig:regolith_structure}). These parameter values approximate the dielectric constants inferred for the shallow subsurface at the Apollo 17 landing site from the inversely modeled profile of \cite{GRIMM2018389}, capturing the range of plausible regolith structures at a well-characterized lunar site. $6 \times 6 \times 6 = 216$ antenna simulations of the far-field beam pattern for 1-50 MHz were created.}
\end{table}

\begin{figure*}
    \centering
    \includegraphics[width=0.9\linewidth]{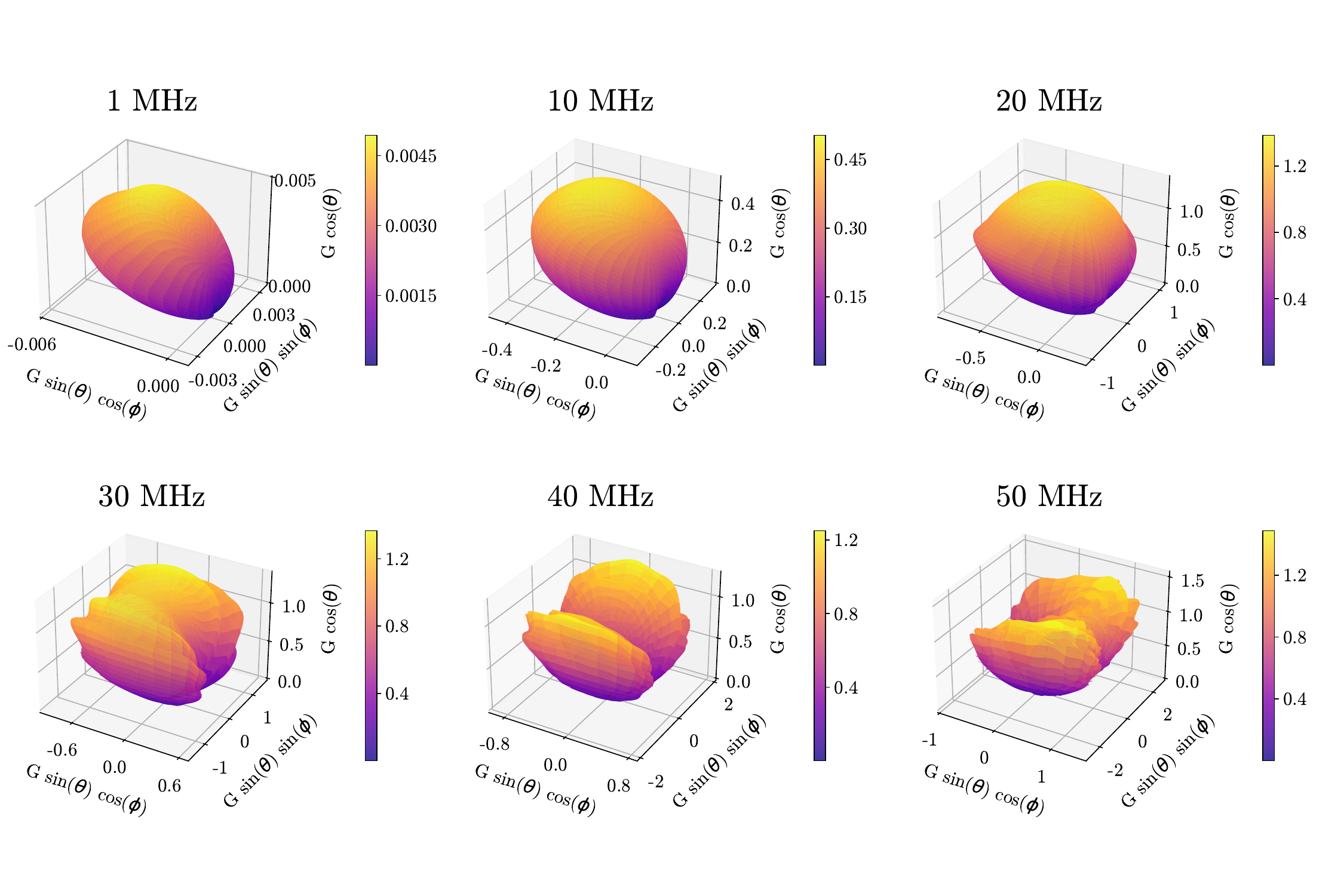}
    \caption{Simulated far-field linear gain patterns of a LuSEE-Night antenna monopole for six frequencies with subsurface parameters $\epsilon_{r, top} = 3.2$, $\epsilon_{r, bottom} = 3.8$, $L=0.5$m. The simulated far-field beam patterns were created with Ansys HFSS, using the model shown in Figure~\ref{fig:figure1}. The surface is plotted in gain-weighted spherical coordinates, where $G$ is the linear gain, $\theta$ is the polar angle measured from the zenith, and $\phi$ is the azimuthal angle.  The gain is computed only over the upper hemisphere, as the subsurface is treated as a boundary condition.}
    \label{fig:gain}
\end{figure*}

\begin{figure}
    \centering
    \includegraphics[width=0.9\linewidth]{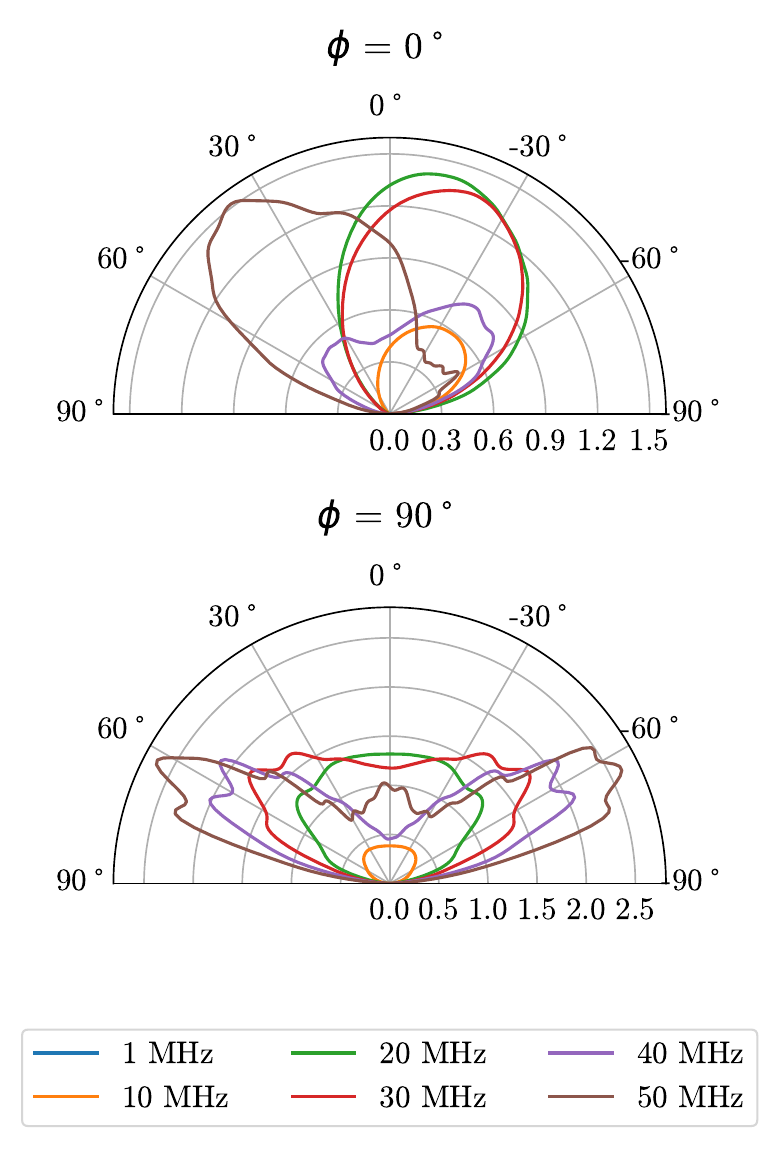}
    \caption{2-D slices of the simulated far-field linear gain of a LuSEE-Night antenna monopole as a function of elevation angle ($\theta$) for six frequencies with subsurface parameters $\epsilon_{r, top} = 3.2$, $\epsilon_{r, bottom} = 3.8$, $L=0.5$m. [Top] For cuts along azimuthal angle $\phi = 0^{\circ}/180^{\circ}$. [Bottom] For cuts along azimuthal angle $\phi = 90^{\circ}/270^{\circ}$. The simulated far-field beam patterns were created with Ansys HFSS, using the model shown in Figure~\ref{fig:figure1}.}
    \label{fig:2D_gain}
\end{figure}

Separating effects of the lunar subsurface on the LuSEE-Night beam from spectral measurements will be crucial for studying and constructing maps of the Galactic foregrounds, and ultimately separating it from the 21-cm Dark Ages signal. In this paper, we explore using Bayesian inference techniques to jointly estimate properties of the lunar subsurface and the Galactic foregrounds with simulated LuSEE-Night voltage power spectral density (PSD) measurements. We simulate LuSEE-Night PSD measurements while varying properties of the lunar subsurface and Galactic foregrounds. We find that Monte Carlo methods can jointly estimate parameters of the lunar subsurface and spectral parameters of the Galactic foregrounds. This means that LuSEE-Night may be able to place constraints on both the dielectric properties of its landing site and the spectral parameters of the low-frequency Galactic foregrounds. 

\section{Analysis} \label{sec:style}

We simulate the time-averaged voltage power spectral density that LuSEE-Night will measure from the lunar far side in the frequency range 1–50 MHz with a frequency resolution of 1 MHz. We simulate spectral measurements for various parameter values of the lunar subsurface and the Galactic foregrounds and construct an emulator function for the LuSEE-Night temperature spectrum. We then use Monte Carlo methods to estimate parameters of the lunar subsurface and the Galactic foregrounds for a simulated measurement dataset.

In this section, we describe our analysis pipeline. We first present the simulated beam models (Section~\ref{sec:beams}) and the Galactic foreground model (Section~\ref{sec:foreground_maps}), followed by the construction of simulated LuSEE-Night measurements (Section~\ref{sec:simulated_measurements}). We then analyze the effect of variations in the structure of the lunar subsurface and in the Galactic foreground parameters on the chromatic behavior of simulated LuSEE-Night measurements via eigenmodes (Section~\ref{sec:eigenmodes}) and perform a joint estimation of foreground and subsurface parameters on a simulated LuSEE-Night measurement (Section~\ref{sec:model_fitting}).

\subsection{Simulated Far-Field Beam} \label{sec:beams}

We simulate the far-field beam pattern of the LuSEE-Night antennas for 216 different 2-layer models of the lunar subsurface. The simulated far-field beam patterns were obtained from electromagnetic simulations with Ansys HFSS \citep{ansys_hfss}. A model of the LuSEE-Night antenna and lander system in HFSS is shown in Figure~\ref{fig:figure1}. The model includes 4 antenna monopole elements labeled North, South, East, and West, which are arranged in a cross-dipole configuration. The 216 far-field beam pattern simulations were created with every permutation and combination of the parameters listed in Table~\ref{tab:regolith_params}.

We simulate the effects of the lunar subsurface by adding layered-impedance boundary conditions to the Ansys HFSS model. The gain is computed only over the upper hemisphere, as the subsurface is treated as a boundary condition. The subsurface is modeled in HFSS as consisting of two layers with dielectric constants $\epsilon_{r1}$ and $\epsilon_{r2}$, respectively, as shown in Figure~\ref{fig:figure2}. The depth of the interface between the two layers $L$ also varies. Modeling the lunar subsurface as a multi-layered stratified medium is consistent with previous work, including \cite{Fa2007-qi}, \cite{Chen2013-zu}, and \cite{GRIMM2018389}. However, this treatment is intended as a simplified demonstration and does not fully represent the expected complexity of the lunar regolith and subsurface structure, which would likely require modeling more layers. 

The values of the 3 parameters, $\epsilon_{r1}$, $\epsilon_{r2}$, and $L$ for each of the 216 models of the lunar subsurface are described in Table~\ref{tab:regolith_params}. The sampled dielectric constants that model the lunar subsurface, as described in Table~\ref{tab:regolith_params}, were chosen because they approximate the top 2-3 layers of the inversely modeled dielectric structure of the Apollo 17 landing site published in \cite{GRIMM2018389}. The dielectric loss tangent of both layers of the lunar subsurface is 0.01. These parameter values are linearly spaced, creating a $6 \times 6 \times 6$ grid of simulated LuSEE-Night beam patterns. 

While this grid is sufficient to capture the broad trends in how subsurface properties affect the antenna response, it is not dense enough to enable reliable interpolation at arbitrary points in parameter space, as discussed in Appendix~\ref{sec:appendix}. As a consequence, the joint inference results presented in Section 2.5 are reliable only when the true parameter values lie on or very close to a grid point, and should be interpreted as a proof-of-concept demonstration rather than a fully validated inference framework. Denser sampling strategies, such as Latin hypercube sampling, are left to future work (Section~\ref{sec:conclusions}).

We use HFSS's Finite Element Method (FEM) solver to simulate the gain, impedance, and electromagnetic fields of the LuSEE-Night antenna-lander system at frequencies between 1 and 50 MHz in 1 MHz steps, assuming perfectly matched layer (PML) boundary conditions. PML boundary conditions simulate an infinitely extending 3D space by absorbing outgoing waves at the edges of the finite simulation volume and preventing artificial reflections. We plot the simulated gain at several frequencies in Figures~\ref{fig:gain} and~\ref{fig:2D_gain}. At frequencies greater than 25 MHz, the antenna's resonance causes the beam pattern to bifurcate. The validation of simulated electromagnetic models of LuSEE-Night and the lunar subsurface will be discussed in forthcoming instrument papers.

\begin{figure}
    \centering
    \includegraphics[width=0.9\linewidth]{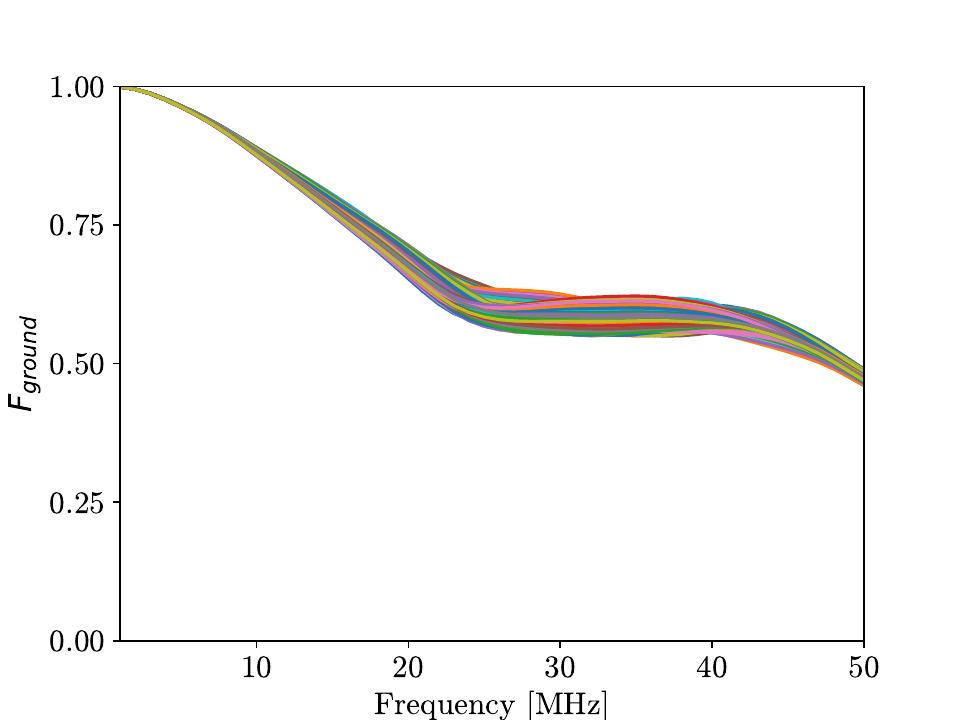}
    \caption{Ground fraction of the simulated LuSEE-Night beam patterns over frequency. The ground fraction is calculated using Equation~\ref{eqn: ground_fraction} and simulated LuSEE-Night antenna far-field beams for a range of values of the two dielectric regolith layers. The parameter values used to simulate these subsurface layers are described in Figure~\ref{fig:regolith_structure} and Table~\ref{tab:regolith_params}.}
    \label{fig:ground_fraction}
\end{figure}

\subsubsection{Ground Fraction} \label{sec:ground_fraction}

We calculate what fraction of each of the simulated LuSEE-Night antenna beam patterns is coupled to the ground, or the ground fraction, using 

\begin{equation}
\label{eqn: ground_fraction}
F_{gnd} (\nu) = 1 - \frac{\iint_{\rm sky}B(\theta, \phi, \nu) d\Omega}{4\pi},
\end{equation}

\noindent where $B(\theta, \phi)$ is the simulated gain of the LuSEE-Night antennas, $\nu$ is frequency, and $\theta$,$\phi$ are spatial coordinates. 

We plot the ground fraction of all 216 simulated beam patterns for 50 frequencies between 1-50 MHz in Figure~\ref{fig:ground_fraction}. We find that, for all frequencies, a significant portion of the LuSEE-Night antenna beam is pointed at the lunar subsurface. Thus, LuSEE-Night will be strongly sensitive to reflections from the lunar subsurface. The spread in ground fraction values across the 216 simulated subsurface models illustrates that the precise dielectric structure of the regolith meaningfully changes how much power is coupled to the ground, motivating the need to jointly constrain subsurface and foreground parameters.

\begin{figure}
    \centering
    \includegraphics[width=0.9\linewidth]{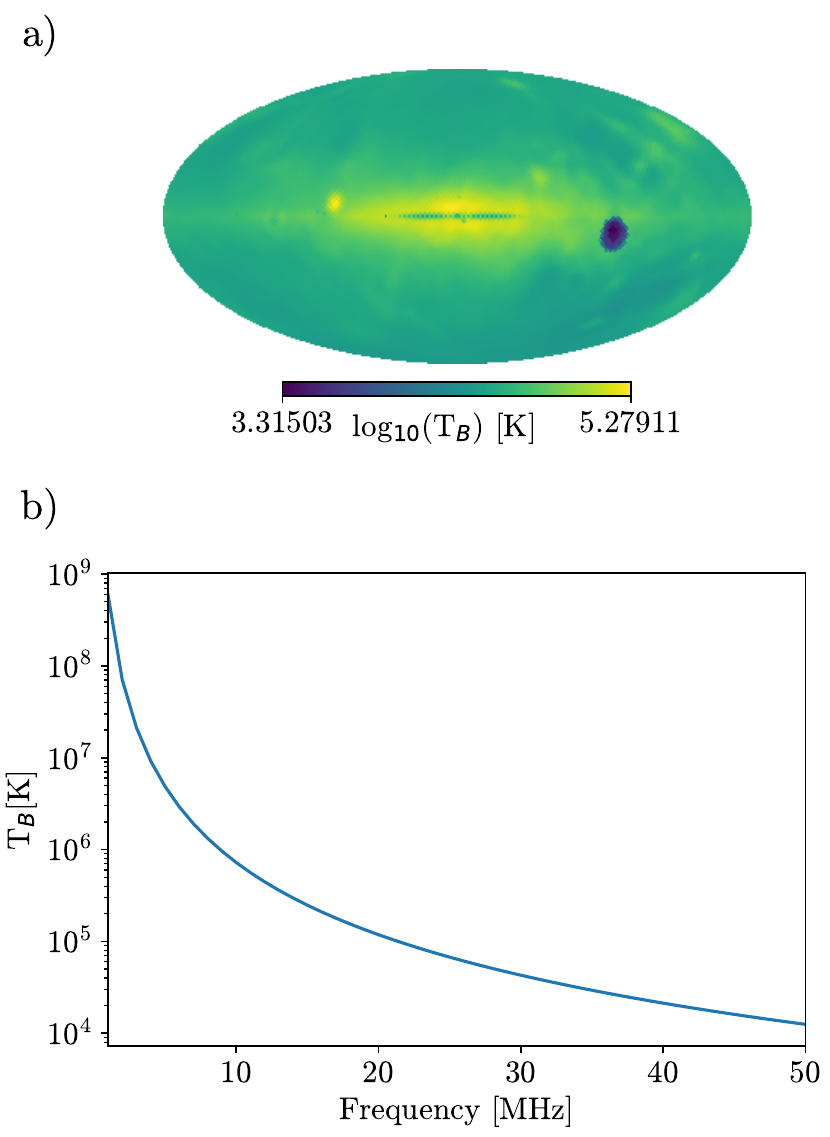}
    \caption{Model of the brightness temperature of the Galactic foregrounds at $\nu = 25$ MHz for $\beta = -2.5$, $\gamma = 0.1$, $k_1 = 1.1$, $k_2 = -10$ as described in Equation \ref{eqn:foreground} (top). Brightness temperature of the pixel of the Galactic foreground map at $l = 270^\circ$, $b = 0^\circ$ over frequency $\nu$ (bottom). These Galactic foreground maps are convolved with the antenna beam patterns according to Equation~\ref{eqn:t_obs_1} to simulate LuSEE-Night measurements.}
    \label{fig:foregrounds2}
\end{figure}

\begin{figure*}
    \centering
    \includegraphics[width=\linewidth]{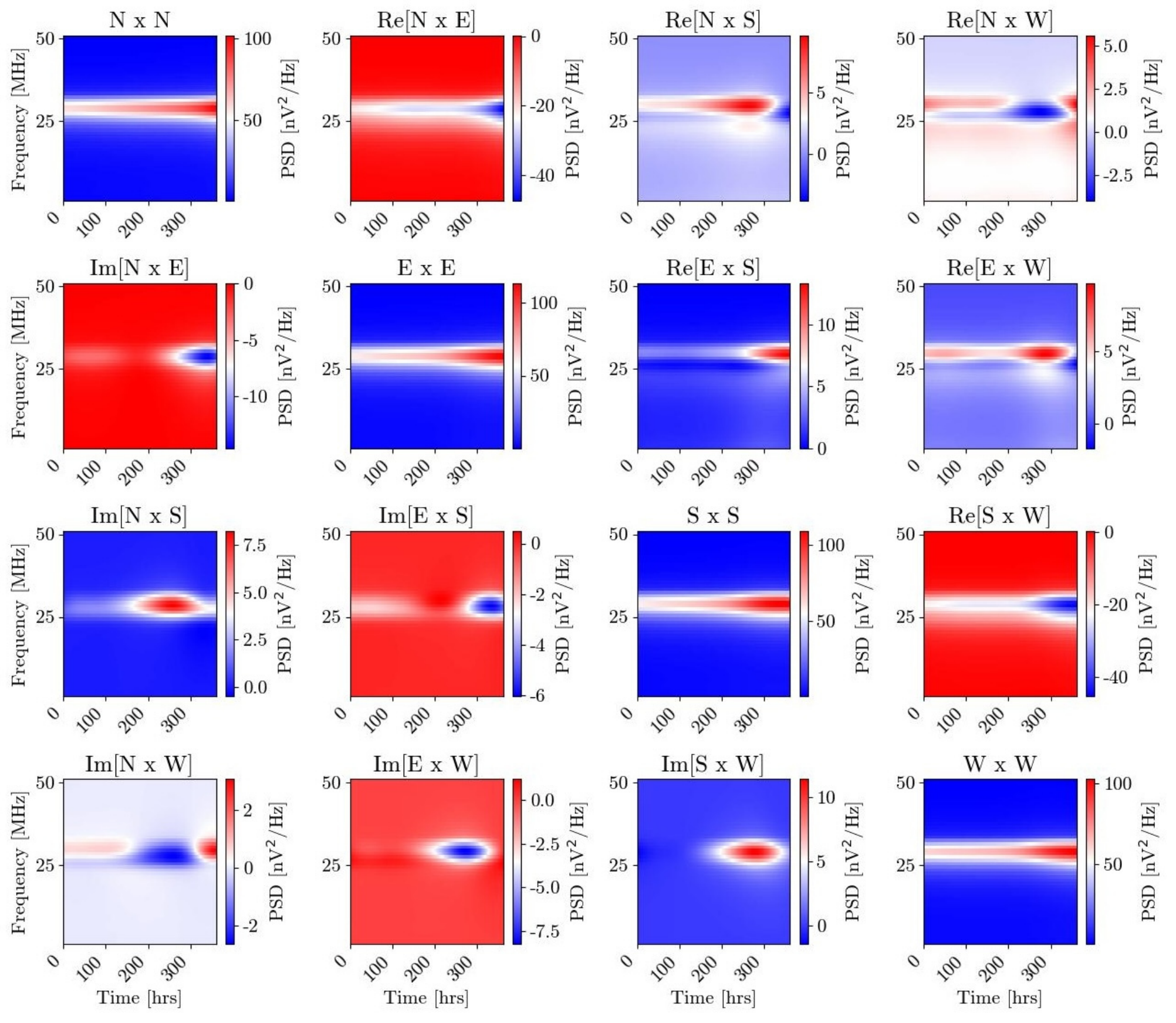}
    \caption{Simulated LuSEE-Night voltage power spectral density (PSD) measurements for the LuSEE-Night antenna monopoles. Each panel shows the PSD as a function of frequency and time for auto-correlations (diagonal) and cross-correlations (off-diagonal) between the four monopoles (N, E, S, W), with real and imaginary components. The PSDs are derived from simulated correlated antenna temperature measurements using Equation~\ref{eqn:t2v2}. These simulated temperature measurements were created with \textsc{luseepy} by convolving a simulated LuSEE-Night beam pattern with a simulated map of the Galactic foregrounds according to Equation~\ref{eqn:t_obs_1}. The LuSEE-Night beam pattern was created using the following subsurface parameters (see Figure~\ref{fig:regolith_structure}): $L=2$m, $\epsilon_{{r,top}} = 3.8$, $\epsilon_{{r,bottom}} = 4.0$. The simulated map of the Galactic foregrounds were created with the following foreground parameters (see Equation~\ref{eqn:foreground} and Figure~\ref{fig:foregrounds2}): $\beta=-2.6$, $\gamma=-0.1$, $k_1=1.1$, $k_2=-90$ K.}
    \label{fig:mock_data_psd}
\end{figure*}

\subsection{Galactic Foreground Maps} \label{sec:foreground_maps}

Following \cite{mozden2016, mozden2018} and \cite{Monsalve_2021}, we model the Galactic foregrounds as varying according to a power law

\begin{equation}
\label{eqn:foreground}
\begin{split}
T_{sky}(\theta, \phi, \nu) = \left[ k_1 T_{\nu_{0}} (\theta, \phi) + k_2 - T_{CMB} \right] \times \\ 
\left( \frac{\nu}{\nu_{0}} \right) ^ {\beta + \gamma \log (\frac{\nu}{\nu_{0}})} + T_{CMB},
\end{split}
\end{equation}

\noindent where $\beta$, $\gamma$, $k_1$, and $k_2$ are frequency- and space-independent fit parameters, $\nu$ is frequency, $\theta$ and $\phi$ are spatial coordinates in the Galactic frame, $T_{\nu_{0}}$ is a reference map, for which we take the Ultralong-wavelength Sky Model with Absorption (ULSA) sky map \citep{Cong_2021-ULSA} at $\nu_0$, $\nu_0$ is the reference frequency of 25 MHz, and $T_{CMB}$ is the CMB temperature 2.725K \citep{Mather_1999}. The Galactic foreground model in Equation~\ref{eqn:foreground} is motivated by \cite{De_Oliveira-Costa2008-lv}, who found that fitting the sky brightness temperature as a polynomial in $\log\left(\frac{\nu}{\nu_0}\right)$, rather than $\nu$ directly, yields more accurate results. An example map of the Galactic foregrounds is shown in Figure~\ref{fig:foregrounds2}. The assumption that the Galactic foregrounds follow a smooth power law-like spectra likely breaks down at $\le10$ MHz, where free-free absorption effects become non-negligible, especially in the direction of the galactic center and plane \citep{1978ApJ...221..114N, 2022A&A...668A.127P}. Despite this limitation, we adopt this model as an initial exploration of the feasibility of jointly constraining sky and subsurface parameters simultaneously; a more physically complete foreground treatment is left to future work. 

\subsection{Simulated LuSEE-Night Voltage Power Spectral Measurements} \label{sec:simulated_measurements}

We simulate the correlated voltage power spectra of a LuSEE-Night monopole on the lunar far side using \textsc{luseepy}\footnote{\url{https://github.com/lusee-night/luseepy}}, a set of python utilities developed for performing various LuSEE-Night related calculations. \textsc{luseepy} utilizes the \textsc{lunarsky}\footnote{\url{https://github.com/aelanman/lunarsky}} package to accurately determine the orientation of the Moon, and thus the instrument, with respect to the celestial sphere at any given time. We first compute the temperature spectrum as a convolution of the antenna beam pattern and the Galactic foregrounds \citep[see e.g.][]{Monsalve2024} 

\begin{equation}
T_{obs}(\nu, t) = \frac{ \\\int_{\rm sky} T_{sky}(\nu, t, \theta, \phi) B(\theta,\phi, \nu) d\Omega}{\int B(\theta,\phi, \nu)  d\Omega}.
\label{eqn:t_obs_1}
\end{equation}

Equation \ref{eqn:t_obs_1} is implemented in \textsc{luseepy} in harmonic space.  We compute $T_{obs}$ for $l_{max}=32$ over one lunar night ($\sim$2 weeks), assuming that the temperature of lunar ground is $0^\circ$K. This assumption is equivalent to assuming that the effect on observations of a realistic moon brightness temperature has been perfectly calibrated out. We will study this effect in future papers. We repeat the calculation for every one of the 216 beam patterns outputted by Ansys HFSS for varying dielectric profiles of the lunar subsurface as described in Section 2.1.

We then convert the simulated observed antenna temperatures to simulated power spectral density (PSD) measurements using

\begin{equation}
\text{PSD}(\nu,t) = 4k_BT_{obs}(\nu,t)\text{Re}[Z_{ant}(\nu)]\gamma^2,
\label{eqn:t2v2}
\end{equation}

\noindent where $k_B$ is the Boltzmann constant, $Z_{ant}$ is the impedance of the LuSEE-Night antenna, $\gamma = \frac{Z_{rec}}{|Z_{ant}+Z_{rec}|}$ is a coupling factor, $Z_{rec}=\frac{1}{i\omega C + \frac{1}{R}}$ is the impedance of the receiver circuit, $C=0$F is the front-end capacitance, and $R=10^6 \Omega$ is the front-end resistance. The antenna impedance $Z_{ant}$ comes from Ansys HFSS simulations for varying dielectric profiles of the lunar subsurface.

The presence and properties of the subsurface affect both the beam pattern and the antenna impedance. Since the antenna impedance enters in the relation between antenna temperature and the observed voltage power spectral density, we \emph{cannot distinguish between the two effects}. 

Figure~\ref{fig:mock_data_psd} plots an example of the simulated voltage power spectral densities of each of the four LuSEE-Night monopoles for a given set of subsurface and foreground parameters. While we assume a smooth power law form for the sky spectrum (an approximation that likely breaks down below $\sim10$ MHz due to free–free absorption), the dominant spectral structure in the simulated LuSEE-Night signal arises near the $\sim30$ MHz resonance peak (Figure~\ref{fig:mock_data_psd}). Consequently, even if the Galactic foreground model is imperfect, the distinctive chromatic structure introduced by the subsurface should remain identifiable, enabling joint constraints on both the lunar dielectric properties and the spectral structure of the galaxy. In Section~\ref{sec:model_fitting}, we construct an emulator of the voltage power spectral density and we perform joint Bayesian inference of subsurface and foreground parameters.

\begin{figure}
    \centering
    \includegraphics[width=\linewidth]{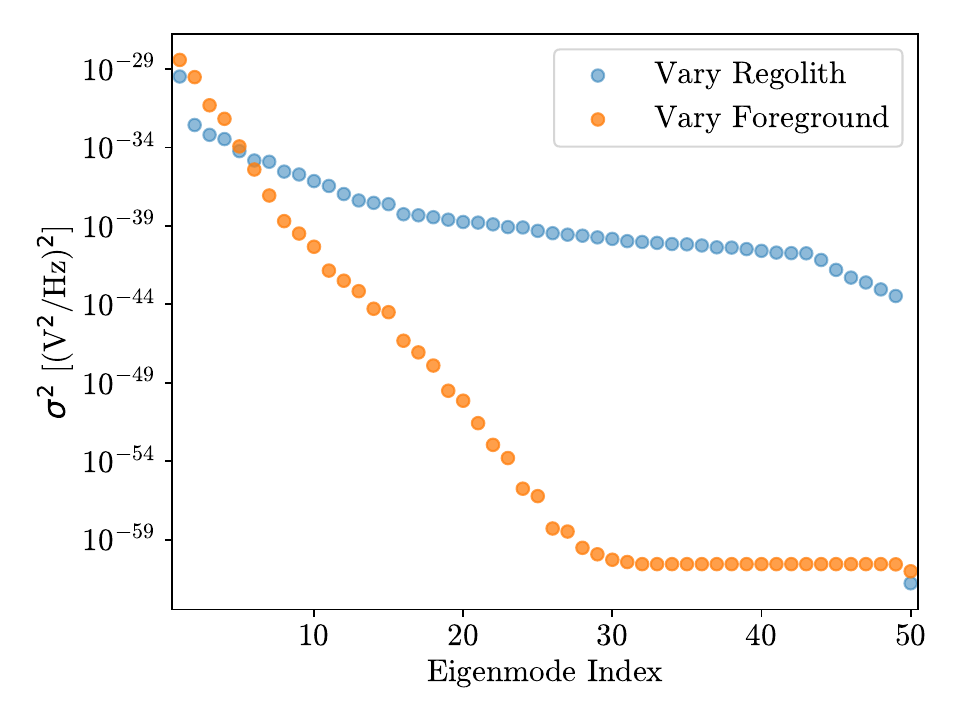}
    \caption{Eigenvalues of the LuSEE-Night voltage power spectral density decompositions for varying subsurface (blue) and foreground (orange) parameters. The rapid decline indicates that the variance in both datasets is dominated by a small number of modes.}
    \label{fig:eigenvalue_spectra}
\end{figure}

\begin{figure*}
    \centering
    \includegraphics[width=\linewidth]{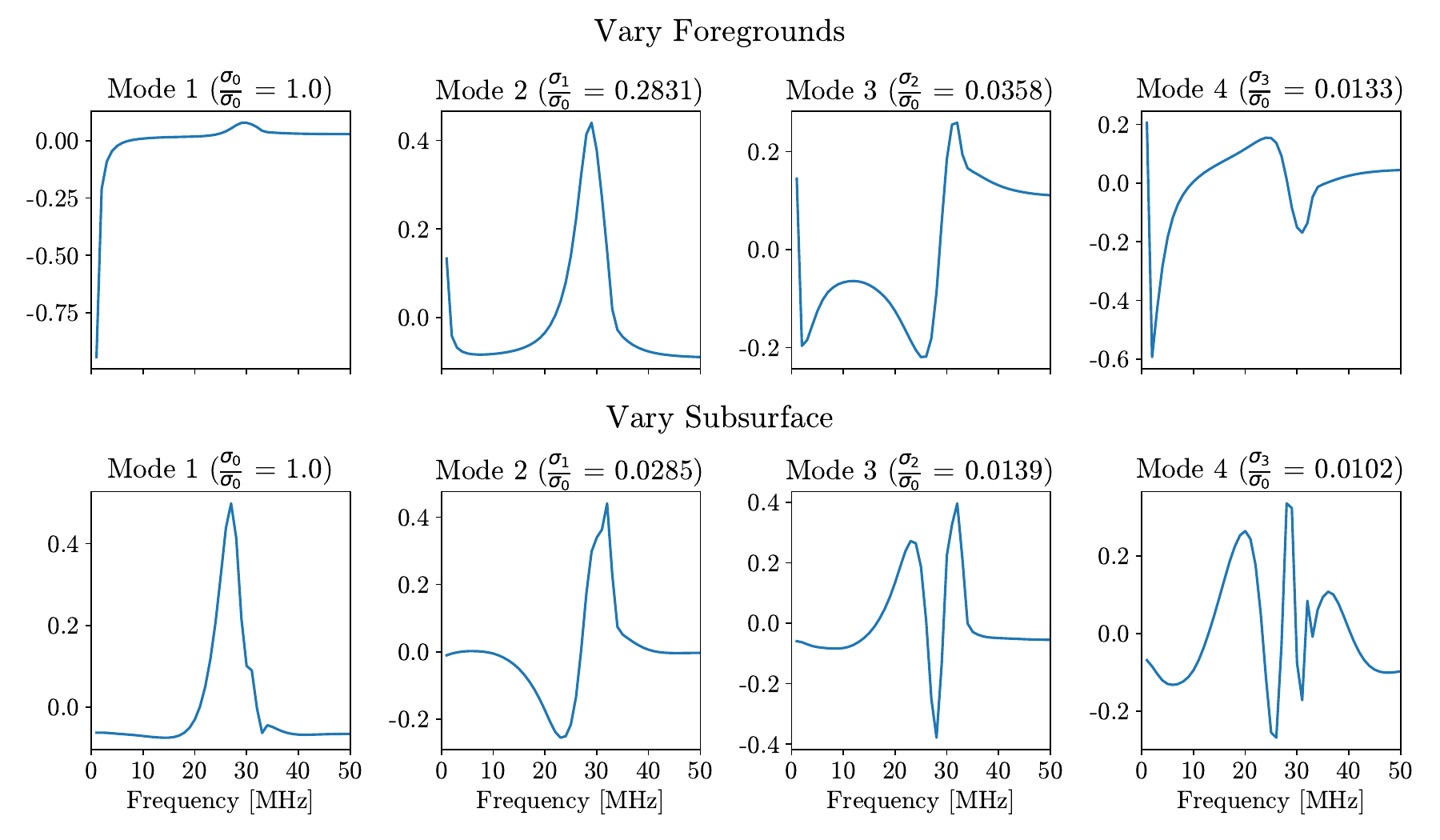}
    \caption{The first four eigenmodes derived from the covariance matrices of the simulated spectra are shown for the subsurface parameter ensemble (bottom) and the foreground parameter ensemble (top). Each eigenmode describes the dominant directions of variation in the simulated power spectra. Foreground modes are smooth and dominated by large spectral slopes at lower frequencies, while subsurface modes exhibit more complex spectral features arising from variations in antenna–surface coupling. The first three eigenvectors for the subsurface effect can be interpreted as affecting the resonance amplitude, shift and width. For each eigenmode, the ratio of the corresponding singular value and the singular value of the first eigenmode is noted in parentheses.}
    \label{fig:eigenmodes}
\end{figure*}

\subsection{Eigenmodes of Simulated LuSEE-Night Measurements with Variations in Subsurface and Foreground Parameters} \label{sec:eigenmodes}

We investigate the effect of variations in the parameters of the lunar subsurface and the Galactic foreground on the simulated measurements separately, in order to characterize the distinct spectral signatures each introduces. Using two separate datasets - one spanning the subsurface parameter space and one spanning the foreground parameter space - each spanning multiple samples of the physical parameters across 50 frequency channels, we construct two ensembles of simulated auto-correlated power spectral densities. To isolate spectral shape variations, each spectrum is mean-subtracted before computing the covariance matrix across the ensemble. We then perform a singular value decomposition (SVD) of the covariance matrix to identify the spectral eigenmodes that describe the dominant directions of variation in the simulated data with respect to the foreground and subsurface parameters.

The eigenvalue spectra (Figure~\ref{fig:eigenvalue_spectra}) show that only a small number of eigenmodes capture the majority of the variance in the simulations for both subsurface-induced and foreground-induced spectral variations. The corresponding eigenmodes (Figure~\ref{fig:eigenmodes}) represent characteristic spectral distortions. Foreground eigenmodes are smooth and dominated by large spectral slopes at lower frequencies and curvature consistent with power law synchrotron emission, whereas subsurface eigenmodes exhibit spectral features arising from changes in antenna–surface coupling and beam response. This is consistent with \cite{rotermund2026} and \cite{tokarsky2022, Tokarsky2023}. \cite{rotermund2026} demonstrates that a horizontal dipole antenna placed close to a dielectric half-space representing the lunar subsurface exhibits strong electrical coupling, significantly modifying the dipole's input impedance and beam pattern.

These results indicate that the main eigenmodes associated with variations in foreground and lunar parameters provide independent information. Therefore, it is justified to expect meaningful constraints from joint fits of foreground and lunar parameters.

\subsection{Model Fitting} \label{sec:model_fitting}

Given the set of simulated LuSEE-Night correlated measurements spanning the parameter space of the lunar subsurface (Table~\ref{tab:regolith_params}) and Galactic foregrounds, we construct an emulator of the LuSEE-Night correlated measurements. An emulator is a surrogate model that approximates the output of a computationally expensive simulation\textemdash in this case, the full LuSEE-Night beam and sky convolution pipeline. Rather than running a new HFSS simulation and sky convolution for every parameter vector proposed by the sampler, the emulator provides a computationally cheap approximation of the LuSEE-Night voltage power spectral density at any point in parameter space. The emulator is built by interpolating between simulated LuSEE-Night measurements across this parameter space. We use Radial Basis Functions (RBF) for this interpolation, as they are well-suited to scattered, multi-dimensional data.

We use this emulator to fit to a mock LuSEE-Night spectral measurement for a set of parameter values of the subsurface and Galactic foregrounds, as indicated in the caption of Figure~\ref{fig:mock_data_psd} and in Table~\ref{tab:results}. We average all the correlated voltage PSD measurements over one lunar night, or 360 hours Local Sidereal Time (LST). We then use a Markov Chain Monte Carlo algorithm to find the parameter vector ($\epsilon_{r,top}, \epsilon_{r,bottom}, L, \beta, \gamma, k_1, k_2$) that maximizes the likelihood function, assuming that the spatial structure of the sky is perfectly known.

\begin{table}
\centering
\caption{Bounds of Uniform Priors for Foreground Parameters
\label{tab:foreground_priors}}
\hspace{-1.5cm}
\begin{tabular}{|c|c|c|}
\hline
{Parameter} & Min & Max \\
\hline
$\beta$ & -3 & -2 \\
$\gamma$ & -1 & +1 \\
$k_1$ & 0 & 2 \\
$k_2$ & -1000 K & +1000 K \\
\hline
\end{tabular}

\tablecomments{Uniform prior bounds for the four Galactic foreground parameters sampled by the PMC algorithm. The priors for the spectral index $\beta$ and curvature $\gamma$ are bounded to physically motivated ranges consistent with low-frequency synchrotron emission \citep{mozden2016, mozden2018, Spinelli2021-le}. The amplitude scaling $k_1$ and offset $k_2$ are given wide bounds to allow flexibility in fitting the foreground spectrum.}
\end{table}

We implement the Markov Chain algorithm with \textsc{pocoMC}, which uses the Preconditioned Monte Carlo (PMC) algorithm \citep{karamanis2022accelerating, karamanis2022pocomc}. We set up the sampler to use a Gaussian log-likelihood function

\begin{equation}
\label{eqn:likelihood}
\log \mathcal{L}  = -\frac{1}{2} \sum_{\nu} \frac{y-y_{emulator}(\theta)}{\sigma_{\nu}^2},
\end{equation}

\noindent where $y$ is the simulated time-averaged LuSEE-Night measurement with 5\% noise added, $y_{emulator}$ is the emulator of the LuSEE-Night voltage power spectral density evaluated at $\theta$, $\theta$ defines the 7 parameters of the Galactic foregrounds and lunar subsurface sampled by the PMC algorithm, and $\sigma_\nu$ represents an assumed Gaussian error that is 5\% of $y$.

We initialize the sampler with uniform prior distributions for all three parameters that model the subsurface and four parameters that model the Galactic foregrounds. The bounds for the priors on the subsurface parameters are described in Table~\ref{tab:regolith_params}. The bounds of the uniform priors for the foreground parameters are described in Table~\ref{tab:foreground_priors}. We run the sampler for 10,000 iterations.

\section{Results}

\begin{figure}
    \centering
    \includegraphics[width=\linewidth]{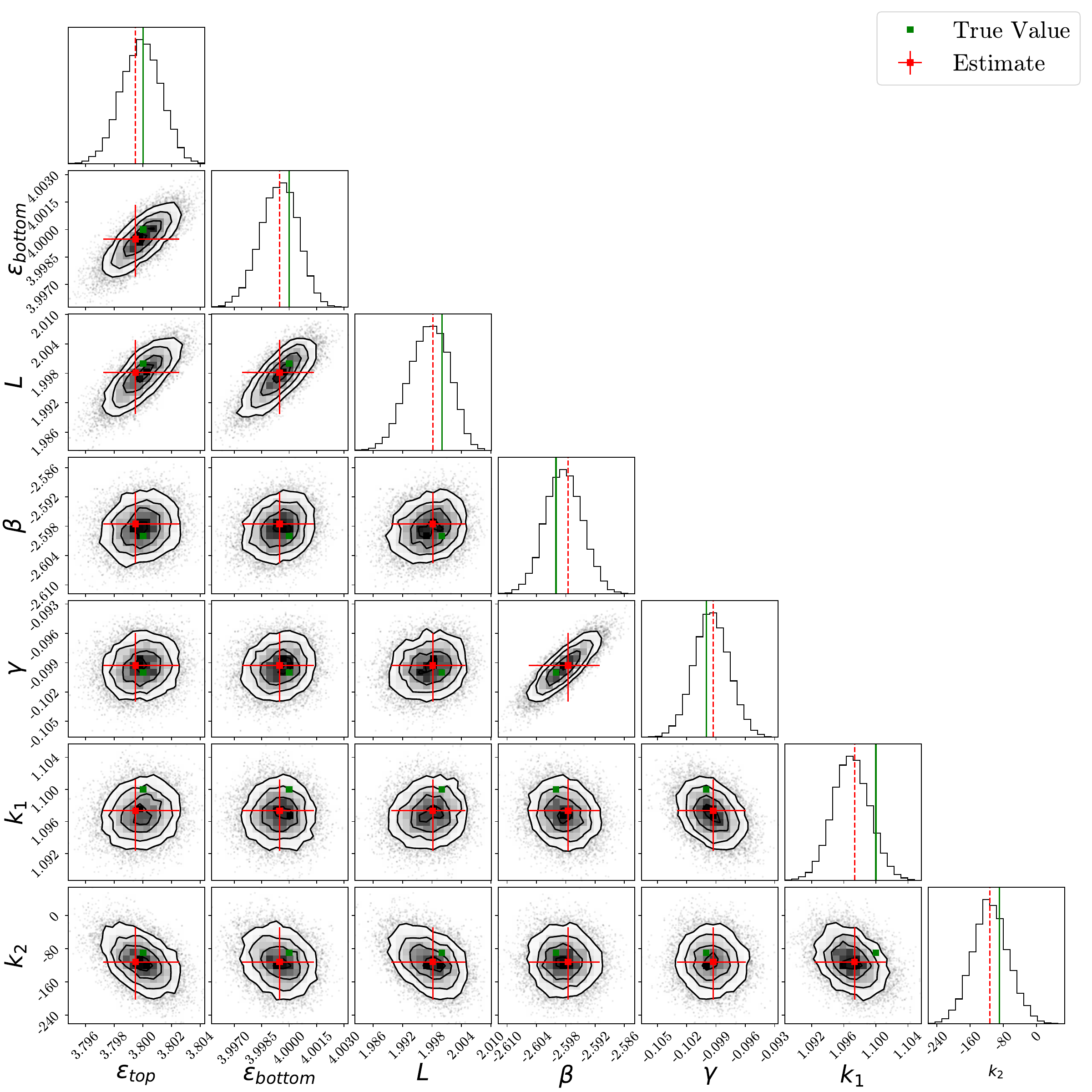}
    \caption{Posterior distribution from fitting a model that emulates LuSEE-Night measurements to a time-averaged mock measurement (Figure~\ref{fig:mock_data_psd}). Parameters that maximize the posterior probability density function (PDF) are plotted in red with 95\% HDI error bars. The parameter values used to generate the mock LuSEE-Night measurement that the model fits to are plotted in green.}
    \label{fig:corner_plot}
\end{figure}

The posterior distributions for the seven model parameters ($\epsilon_{r, top}, \epsilon_{r, bottom}, L, \beta, \gamma, k_1, k_2$) sampled by the PMC algorithm are shown and described in Figure~\ref{fig:corner_plot} and Table~\ref{tab:results}. We find that the PMC algorithm converges to a posterior distribution which comfortably encompasses the true input parameters.

\begin{figure}
    \centering
    \includegraphics[width=\linewidth]{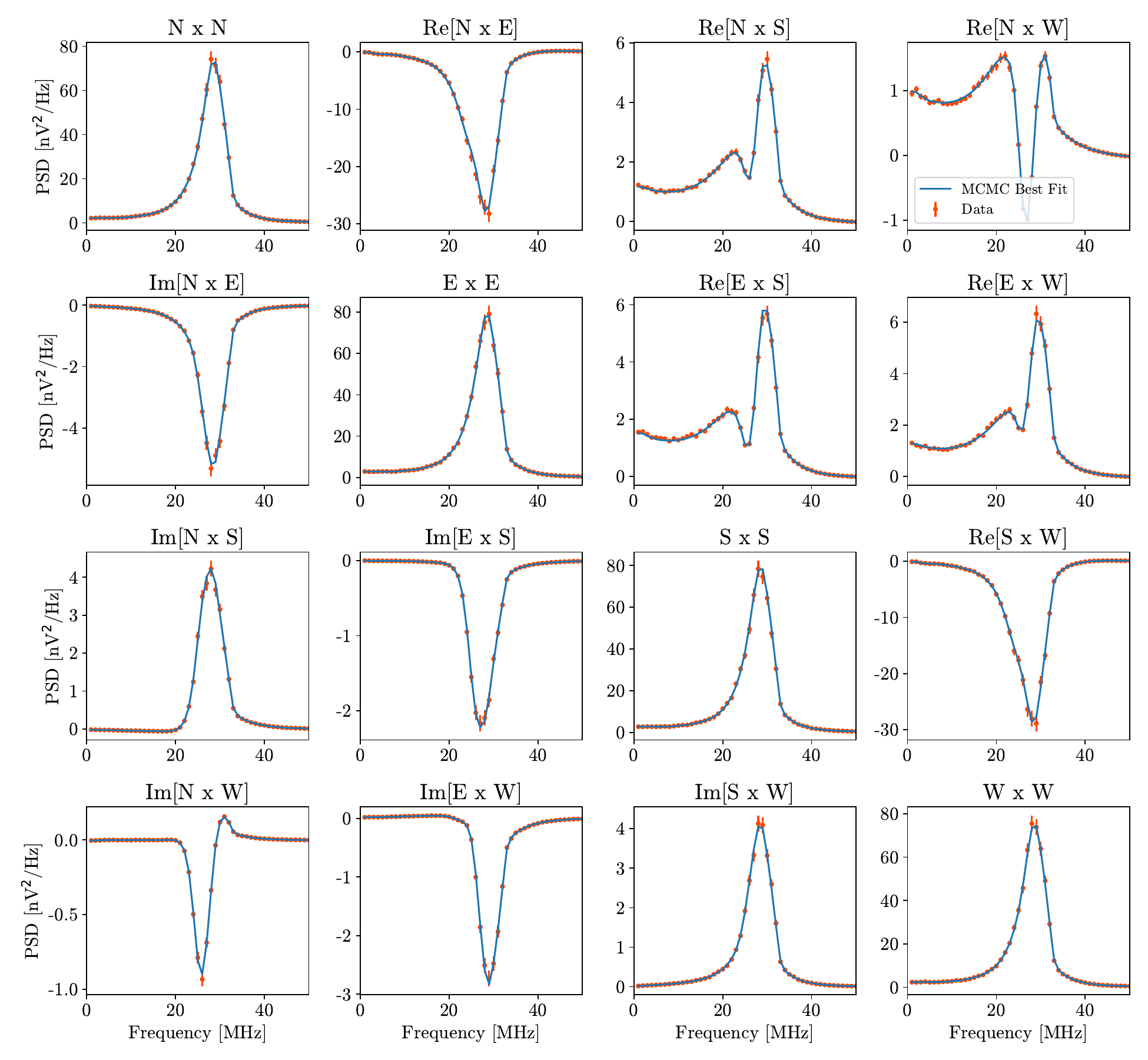}
    \caption{Comparing the PMC fit (Figure~\ref{fig:corner_plot}) with the time-averaged mock LuSEE-Night measurement (Figure~\ref{fig:mock_data_psd}). The red data points are the simulated time-averaged power spectral densities, which were used as a mock LuSEE-Night measurement to fit the model to. The red error bars represent an assumed 5\% Gaussian error $\sigma_\nu$. The solid blue line is the output of the emulator evaluated at the maximum a posteriori parameter vector}
    \label{fig:posterior}
\end{figure}

Although correlations are clearly identifiable in the joint posterior distributions shown in Figure~\ref{fig:corner_plot}, all of the parameters that describe the subsurface and foregrounds are well-constrained within the assumed settings. The posteriors for these parameters are approximately Gaussian, with 95\% high density intervals noted in Table~\ref{tab:results}.

The subsurface parameters ($\epsilon_{r, top}$, $\epsilon_{r, bottom}$, $L$) exhibit moderate positive correlations, as each contributes to shaping the antenna–surface coupling and the resonance structure of the measured spectrum. Similarly, the foreground spectral parameters, $\beta$ and $\gamma$, are positively correlated, since both influence the slope and curvature of the foreground spectrum.

Despite these correlations, the overall parameter space remains well constrained. As shown in Section~\ref{sec:eigenmodes}, foreground-induced variations are dominated by lower-frequency structure, whereas subsurface-induced variations introduce higher-frequency structure associated with the antenna resonance. The independent information provided by these spectral components enables the Bayesian framework to disentangle the two effects and jointly constrain all parameters.

In Figure~\ref{fig:posterior}, we present a posterior predictive check and compare the fit to the mock LuSEE-Night measurement. The model provides an excellent fit across all 16 correlated spectral products, with residuals consistent with the assumed 5\% Gaussian noise level. This agreement demonstrates that the inference framework can, in principle, recover these parameters without significant bias. 

However, we note that this level of precision represents an optimistic scenario, as several simplifying assumptions separate our simulated analysis from a real LuSEE-Night observation. First, the foreground model adopted here (Equation~\ref{eqn:foreground}) assumes a smooth power law spectral form across the full 1–50 MHz band. In reality, we expect that the sky spectrum deviates from a simple power law at frequencies below $\sim$10 MHz due to free-free absorption along the Galactic plane \citep{1978ApJ...221..114N, 2022A&A...668A.127P}, and the true spatial and spectral structure of the low-frequency sky remains poorly constrained at these frequencies \citep{De_Oliveira-Costa2008-lv, Cong_2021-ULSA}. Second, the beam simulations assume a simplified two-layer subsurface model with a small number of free parameters, whereas the true lunar regolith is expected to exhibit a more complex, multi-layered dielectric structure \citep{GRIMM2018389, Fa2007-qi}. Third, the assumed 5\% Gaussian noise level is idealized. Real LuSEE-Night measurements will be subject to receiver noise with frequency-dependent characteristics and systematic instrumental effects, all of which could broaden the posterior distributions or introduce biases not captured in this analysis. Fourth, we note that the mock data used for inference are drawn from the same foreground model family used in the forward model, meaning that the recovery demonstrates internal consistency rather than robustness to foreground model misspecification. Testing the pipeline against mock data generated with an independent foreground model is a validation step that we defer to future work, alongside the development of a denser simulation grid and a more physically complete foreground treatment (Section~\ref{sec:conclusions}).

\begin{table}
\centering
\caption{Evaluating Posterior Distribution Outputted by Markov Chain Algorithm
\label{tab:results}}
\begin{tabular}{|c|c|c|c|}
\hline
{Parameter} & $\theta_{data}$ & $\theta_{rec}$ & 95\% HDI \\
\hline
$\epsilon_{r, top}$ & 3.8 & 3.7995 & 3.7972 - 3.8025 \\
$\epsilon_{r, bottom}$ & 4.0 & 3.9995 & 3.9974 - 4.0014 \\
$L$ & 2.0 m & 1.9981~m & 1.9896 - 2.0048 m \\
$\beta$ & -2.6 & -2.5975 & -2.6056 - -2.5911 \\
$\gamma$ & -0.1 & -0.0993 & -0.1030 - -0.0960 \\
$k_1$ & 1.1 & 1.0973 & 1.0923 - 1.1012 \\
$k_2$ & -90 K & -112.1601 K & -202.67 - -29.101 K \\
\hline
\end{tabular}

\tablecomments{$\theta_{data}$ is the vector of subsurface and Galactic foreground parameter values used to create the mock LuSEE-Night measurement. $\theta_{rec}$ is the recovered posterior. We also note the 95\% High Density Interval (HDI) of the posterior distribution.}
\end{table}

In the example shown above, the fiducial model used to generate data was chosen to lie on the parameter grid. When repeated with off-grid parameters, results are often unstable, indicating that our model grid is not sufficiently dense to enable reliable interpolation across parameters.

\section{Conclusions and Future Work} \label{sec:conclusions}

We demonstrate that Bayesian inference methods can jointly estimate properties of the lunar subsurface and spectral parameters of the Galactic foregrounds from simulated LuSEE-Night correlated voltage power spectral density measurements. The joint fitting framework exploits the approximate orthogonality of the spectral signatures introduced by the subsurface and the foregrounds: subsurface effects are localized around the antenna resonance near 30 MHz, while foreground variations produce smooth, broadband spectral structure. 

Simulations of LuSEE-Night beam patterns for varying subsurface parameters have shown that LuSEE-Night is highly sensitive to reflections from the subsurface. We simulate 16 LuSEE-Night correlated spectral densities for varying values of parameters modeling the lunar subsurface and Galactic foregrounds. We interpolate between these time-averaged simulated voltage power spectral densities, which are based on the ULSA sky model and HFSS antenna simulations, to construct an emulator that is a function of the parameters modeling the lunar subsurface and Galactic foregrounds. We find that a Preconditioned Monte Carlo (PMC) algorithm with a Gaussian log-likelihood function converges to a posterior distribution that estimates the lunar subsurface and foreground spectral parameters to within 5\%.

The main conclusions of this paper can be summarized as follows:
\begin{itemize}

    \item Effects of the subsurface are localized around the main spectral feature of LuSEE-Night antenna response, which is the resonance associated with the antenna size. The shape of this feature should be constrained in a way that is fairly independent on the assumptions about the sky signals, which cannot know about geometry of LuSEE-Night and are known to be smooth over frequency ranges corresponding to the feature size. Therefore, despite operating with a very simple model, our main result that the subsurface and sky properties can be estimated simultaneously is likely robust.
    
    \item To correctly describe the effects of the subsurface on the antenna, a denser and more sophisticated sampling of the parameter space than used in this paper is required. 
    
    \item Antenna temperature is not an observable for an instrument like LuSEE-Night. Since subsurface affects both the antenna beam pattern and antenna impedances, forecasts need to be performed using observed voltage power spectral density rather than antenna temperature.

\end{itemize}

In the future, we will attempt to model the HFSS antenna beam pattern results using analytical models. These analytical models consider a wave reflection at the subsurface-space boundary and at the impedance layer boundaries to construct an analytical approach to correct antenna response for subsurface properties. Future models will also involve more complex subsurface models with more parameters and less symmetry. In future work, we will employ techniques such as latin hyper-cube in order to increase fidelity while keeping resources required to perform simulations within reasonable bounds. The existing pipeline is well-suited to accommodate denser latin hyper-cube sampling strategies with minimal modification. The emulator is built using Radial Basis Functions (RBF), which are agnostic to the specific structure of the input grid and can interpolate over scattered, irregularly spaced points. Replacing the current $6\times6\times6$ Cartesian grid with Latin hypercube samples therefore requires minimal architectural changes to the emulator or the Bayesian inference pipeline. The primary bottleneck for denser sampling is the computational cost of the HFSS simulations themselves, not the inference framework. How to optimally parameterize and grid such complex model spaces remains an open problem.

\begin{acknowledgments}
This research was supported by the U.S. Department of Energy, Office of Science under the LuSEE-Night Science program. The LuSEE-Night program is funded under NASA Contract 80MSFC22CA018 and DOE Office of Science Contract DE-SC0012704. This work was also supported in part through HPC computational resources provided by NERSC (ERCAP0033114). Stuart D. Bale acknowledges the support of the Royal Society Wolfson Visiting Fellowship program. David Rapetti acknowledges support by NASA APRA grant award 80NSSC23K0013.
\end{acknowledgments}





%
\facilities{NERSC}

\appendix

\section{Evaluating Emulator Accuracy} \label{sec:appendix}

We evaluate how well the emulator described in Section~\ref{sec:model_fitting}, which was constructed by interpolating between simulated LuSEE-Night measurements, approximates the simulated LuSEE-Night measurement at parameter values that were not included in the original $6\times6\times6$ beam pattern simulation grid. This is important because the joint inference pipeline described in Section~\ref{sec:model_fitting} relies on the emulator to predict LuSEE-Night voltage power spectral densities at arbitrary points in parameter space, and interpolation errors could bias or broaden the recovered posterior distributions when fitting to data not on the original simulation grid.

\begin{figure*}
    \centering
    \includegraphics[width=0.8\linewidth]{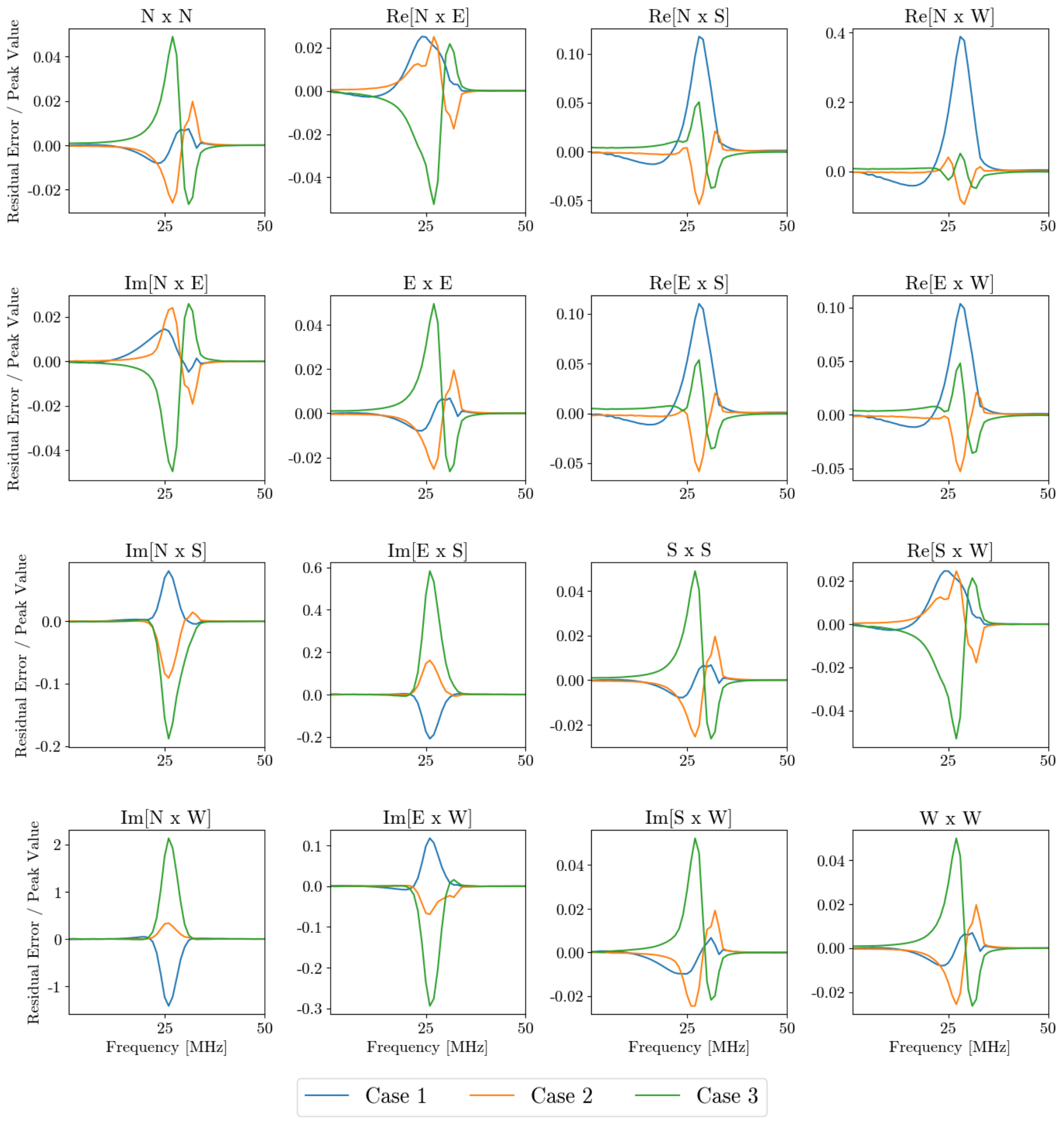}
    \caption{Ratio of the residuals between the emulator and fiducial simulated LuSEE-Night voltage PSD measurements and the maximum value of the simulated LuSEE-Night voltage PSD measurements. We consider three cases of off-grid subsurface parameter vectors described in Table~\ref{tab:emulator_cases}. Each panel shows the difference between the emulator output and the fiducial simulation for all 16 correlated spectral products (auto-correlations on the diagonal, cross-correlations off-diagonal) between the four monopoles (N, E, S, W). Blue, orange, and green lines correspond to Case 1 (one off-grid parameter), Case 2 (two off-grid parameters), and Case 3 (all three parameters off-grid), respectively. The residuals tend to increase as more parameters are moved off the grid, indicating that the $6\times6\times6$ simulation grid is not sufficiently dense to support reliable interpolation at arbitrary points in parameter space.}
    \label{fig:emulator_accuracy}
\end{figure*}

We consider three cases of increasing difficulty, described in Table~\ref{tab:emulator_cases}. In Case 1, only one subsurface parameter ($\epsilon_{r,top}$) is moved off the grid, while the other two ($\epsilon_{r,bottom}$ and $L$) remain at grid values. In Case 2, two parameters ($\epsilon_{r,top}$ and $\epsilon_{r,bottom}$) are off-grid simultaneously. In Case 3, all three subsurface parameters are off-grid. These cases are designed to test how the emulator accuracy degrades as the parameter vector moves further from the nearest grid points.

For each of the three cases described in Table~\ref{tab:emulator_cases}, we simulate the LuSEE-Night far-field beam pattern with Ansys HFSS as described in Section~\ref{sec:beams}. We simulate the time-averaged voltage power spectral density that LuSEE-Night will measure as described in Section~\ref{sec:simulated_measurements}. For all three simulated LuSEE-Night measurements, we fix the Galactic foreground parameters to $\beta=-2.6$, $\gamma = -0.1$, $k_1 = 1.1$, and $k_2 = -90^\circ$K, so that any discrepancy between the emulator and the fiducial simulation is attributable solely to interpolation error in the lunar subsurface parameter space.

\begin{table}[b!]
\centering
\caption{Cases for Evaluating Emulator Accuracy \label{tab:emulator_cases}}
\hspace{-2cm}
\begin{tabular}{ |c|c|c|c| } 
\hline
Case & $\epsilon_{r,top}$ & $\epsilon_{r, bottom}$ & $L$ \\
\hline
1 & 3.642 & 3.8 & 0.5 m \\ 
2 & 3.75 & 3.92 & 3 m \\
3 & 3.571 & 4.693 & 1.887 m \\
\hline
\end{tabular}
\tablecomments{Off-grid subsurface parameter vectors used to evaluate emulator accuracy. In Case 1, only $\epsilon_{r,top}$ is moved off the $6\times6\times6$ simulation grid while $\epsilon_{r,bottom}$ and $L$ remain at grid values. In Case 2, both $\epsilon_{r,top}$ and $\epsilon_{r,bottom}$ are off-grid. In Case 3, all three subsurface parameters are off-grid. For all three cases, the Galactic foreground parameters are fixed at $\beta = -2.6$, $\gamma = -0.1$, $k_1=1.1$, and $k_2=-90$ K. Emulator accuracy for each case is described in Figure~\ref{fig:emulator_accuracy}.}
\end{table}

We then evaluate the emulator described in Section~\ref{sec:model_fitting} at the three off-grid parameter vectors in Table~\ref{tab:emulator_cases}, and compute the residuals between the emulator output and the fiducial simulation for all 16 correlated spectral products. The ratio of these residuals and the maximum value of the fiducial simulation of the LuSEE-Night correlated voltage PSD measurements are plotted in Figure~\ref{fig:emulator_accuracy}.

As shown in Figure~\ref{fig:emulator_accuracy}, the emulator residuals grow as more subsurface parameters are moved off the grid. In Case 1, where only a single parameter is off-grid, the residuals are relatively smaller across most of the band. For all three cases, the residuals are largest near the $\sim$30 MHz antenna resonance where the beam response varies most rapidly with subsurface properties. In Case 3, where all three parameters are simultaneously off-grid, the residuals are largest, and their amplitude is comparable to or exceeds the 5\% noise level assumed in the joint inference, indicating that a denser sampling of parameter space is required for unbiased inference. 

This demonstrates that our inference results can be interpreted as valid in the regime where the emulator error is subdominant to measurement noise. This also motivates the use of denser and more sophisticated sampling strategies in future work, as discussed in Section~\ref{sec:conclusions}.

\bibliography{sample701}{}
\bibliographystyle{aasjournalv7}



\end{document}